\documentclass[useAMS,usenatbib]{mn2e}
\usepackage{graphicx}
\usepackage{color}
\def\yonetoku{$E_{\rm p}$--$L_{\rm p}$ }
\def\amati{$E_{\rm p}$--$E_{\rm iso}$ }

\title[$E_{\rm p}$--$E_{\rm iso}$ and $E_{\rm p}$--$L_{\rm p}$ Correlations for Short Gamma Ray Bursts ]
{Possible existence of $E_{\rm p}$--$L_{\rm p}$ and $E_{\rm p}$--$E_{\rm iso}$ correlations for Short Gamma-Ray Bursts 
with a factor 5 to 100 dimmer than  those for Long Gamma-Ray Bursts}
\author[R. Tsutsui et al.]
{Ryo Tsutsui$^{1}$\thanks
{E-mail: tsutsui@resceu.s.u-tokyo.ac.jp(RT)},
Daisuke Yonetoku$^{2}$
\newauthor
Takashi Nakamura$^{3}$,
Keitaro Takahashi$^{4}$,
and
Yoshiyuki Morihara$^{2}$\\
$^{1}$Research Center for the Early Universe, School of Science, University of Tokyo, 
Bunkyo-ku, Tokyo 113-0033, Japan\\
$^{2}$Department of Physics, Kanazawa University,
Kakuma, Kanazawa, Ishikawa 920-1192, Japan\\
$^{3}$Department of Physics, Kyoto University,
Kyoto 606-8502, Japan \\
$^{4}$Faculty of Science, Kumamoto University,
Kurokami, Kumamoto, 860-8555, Japan}
\begin{document}


\pagerange{\pageref{firstpage}--\pageref{lastpage}} \pubyear{2008}

\maketitle

\label{firstpage}

\begin{abstract}
We analyzed correlations among the rest frame spectral peak energy $E_{\rm p}$, the observed frame 64ms peak isotropic luminosity $L_{\rm p}$ and the isotropic energy $E_{\rm iso}$ for 13 Short Gamma Ray Burst (SGRB) candidates having the measured redshift $z$, $T_{\rm 90}^{\rm obs}/(1+z)<2$ sec and well determined spectral parameters. A SGRB candidate is regarded as a misguided SGRB if it is located in the 3-$\sigma_{\rm int}$ dispersion region from the best-fit function of the $E_{\rm p}$--$E_{\rm iso}$ correlation for Long GRBs (LGRBs) while the others are regarded as secure SGRBs possibly from compact star mergers. Using 8 secure SGRBs out of 13 SGRB candidates, we tested whether $E_{\rm p}$--$E_{\rm iso}$ and $E_{\rm p}$--$L_{\rm p}$  correlations exist for SGRBs. We found  that $E_{\rm p}$--$E_{\rm iso}$ correlation for SGRBs($E_{\rm iso} =10^{51.42 \pm 0.15}{\rm erg~ s^{-1}} ({E_{\rm p}}/{\rm 774.5~keV})^{1.58 \pm 0.28}$) seems to exist  with the correlation coefficeint $r=0.91$ and chance probability $p=1.5\times10^{-3}$. We found also that  the $E_{\rm p}$--$L_{\rm p}$ correlation for SGRBs($L_{\rm p} = 10^{52.29 \pm 0.066}{\rm erg~ s^{-1}} ({E_{\rm p}}/{\rm 774.5~keV})^{1.59 \pm 0.11}$) is tighter than $E_{\rm p}$--$E_{\rm iso}$ correlation since $r=0.98$ and $p=1.5\times10^{-5}$. Both correlations for SGRBs are dimmer than those of LGRBs for the same $E_{\rm p}$ by factors $\sim$100 ($E_{\rm p}$--$E_{\rm iso}$) and $\sim$ 5($E_{\rm p}$--$L_{\rm p}$). Applying the tighter $E_{\rm p}$--$L_{\rm p}$ correlation for SGRBs to 71 bright BATSE SGRBs, we found that pseudo redshift $z$ ranges from $0.097$ to  $2.258$ with the mean $<z>$ of 1.05. The redshifts of SGRBs apparently  cluster at lower redshift than those of LGRBs ($<z>\sim 2.2 $), which supports the merger scenario of SGRBs.
\end{abstract}


\section{Introduction}
\label{sec:introduction}

For Long Gamma Ray Bursts (LGRBs), several observational correlations 
among the rest frame spectral peak energy $E_ {\rm p}$, the peak isotropic 
luminosity $L_{\rm p}$ and the isotropic energy $E_{\rm iso}$
in the prompt emission phase have been proposed. $E_{\rm p}$--$E_{\rm iso}$ correlation was
first reported by \citet{Amati:2002} and argued by many
authors \citep{Sakamoto:2004,Lamb:2004,Amati:2006,Amati:2009,Yonetoku:2010}. 

As for $L_{\rm p}$, \citet{Yonetoku:2004} reported a rather tight correlation 
between $E_{\rm p}$ and the observed frame 1-second peak isotropic luminosity 
$L_{\rm p}$. In 2004, the number of LGRBs with well determined redshifts
and spectral parameters was only 16. Nevertheless, the correlation was
found to be very tight: the linear correlation coefficient ($r$) between
$\log E_{\rm p}$ and $\log L_{\rm p}$ is $0.958$ and the chance probability ($p$)
is $5.3\times 10^{-9}$. Several authors argued on the property of the $E_{\rm p}$--$L_{\rm p}$
correlation \citep{Ghirlanda:2005a,Ghirlanda:2005b, Krimm:2009} and 
confirmed the existence. \citet{Tsutsui:2009b} found that
adding a new observables $T_ {\rm L}$, the luminosity time defined by
$T_{\rm L}=E_{\rm iso}/L_{\rm p}$, improves the correlation and gave
$E_{\rm p}$--$T_{\rm L}$--$L_{\rm p}$ correlation. In $E_{\rm p}$--$T_{\rm L}$--$L_{\rm p}$ correlation, the intrinsic dispersion is
reduced by $\sim$ 40 \% compared with the $E_{\rm p}$--$E_{\rm iso}$ and $E_{\rm p}$--$L_{\rm p}$ correlations.

\citet{Ghirlanda:2004} applied the $E_{\rm p}$--$L_{\rm p}$ correlation to 
bright short Gamma Ray Bursts (SGRBs) observed by BATSE without 
measured redshift. That is, they assumed that SGRBs obey the same
$E_{\rm p}$--$L_{\rm p}$ correlation of LGRBs and estimated the pseudo redshifts
of SGRBs although no evidence for the existence of the $E_{\rm p}$--$L_{\rm p}$ correlation
for SGRBs at that time. They found that the pseudo redshifts are obtained
for all selected SGRBs and the distribution is similar to that of LGRBs
known at that time. On the other hand,
\citet{Nakar:2005,Band:2005,Butler:2007,Shahmoradi:2010} argued that
$E_{\rm p}$--$L_{\rm p}$ correlation might be due to selection effects, since
$E_{\rm p}$ was determined from the time integrated spectra.
However, \citet{Ghirlanda:2010} showed that in the individual 
pulses of several LGRBs, $E_{\rm p}$--$L_{\rm p}$ correlation holds for each pulse
even though $E_ {\rm p}$ changes an order of magnitude from pulse to pulse. 
Similar property was found for GRB061007 by \citet{Ohno:2009}.
These results strongly suggest that $E_{\rm p}$--$L_{\rm p}$ correlation is  not 
a result of selection biases but a real physical one.

As for SGRBs, the number of SGRBs with measured redshifts and
$E_{\rm p}$ was so small that it was difficult to check
if $E_{\rm p}$--$L_{\rm p}$ correlation holds or not. However, \citet{Ghirlanda:2011}
showed that for 14 Fermi/GBM SGRBs without redshifts, the individual pulses
follow a relation of $E_{\rm p}$ $\propto F_{\rm pulse}^s$ with $s\sim 1$
where $F_{\rm pulse}$ is the observed energy flux. This reminds us
what happened to the individual pulses of LGRBs in \citet{Ghirlanda:2010} 
and suggests that a similar correlation might exist even for SGRBs
in the rest frame.

In this study, we select 13 SGRB candidates with well determined
redshift,  spectral parameters, $E_ {\rm p}$, $L_ {\rm p}$ and
$E_{\rm iso}$ to see if the correlations among $E_ {\rm p}$,
$L_ {\rm p}$ and $E_{\rm iso}$ exist. In section 2, we will
show that our criteria on SGRBs yield 8 secure SGRBs out of
13 SGRB candidates. Using these SGRBs, we examine if the $E_{\rm p}$--$E_{\rm iso}$ and $E_{\rm p}$--$L_{\rm p}$
correlations exist or not. In section 3, we will apply the $E_{\rm p}$--$L_{\rm p}$
correlation obtained in section 2 to 71 bright BATSE SGRBs
without measured redshift to determine the pseudo redshift $z$. 
Section 4 will be devoted to discussions. Throughout the paper
we adopt a cosmological model with $\Omega_\Lambda=0.7 $,
$\Omega_m=0.3 $ and $H_0=70{\rm km s^{-1}Mpc^{-1}}$

\section{SGRBs with well determined redshift $z$, $E_ {\rm p}$, $L_ {\rm p}$ and $E_{\rm iso}$}

In the previous works, it has been checked whether SGRBs are consistent with the $E_{\rm p}$--$E_{\rm iso}$ and $E_{\rm p}$--$L_{\rm p}$ correlations for LGRBs. First, \cite{Amati:2006} showed that two short GRBs are clear outliers of the $E_{\rm p}$--$E_{\rm iso}$ correlation. Then, \cite{Ghirlanda:2009} found that their six SGRBs are inconsistent with the $E_{\rm p}$--$E_{\rm iso}$ correlation, while they possibly follow the $E_{\rm p}$--$L_{\rm p}$ correlation. Now, by the end of 2011, there are more than 10 SGRBs which have well-determined redshifts and spectral parameters so that we can check more systematically if SGRBs are consistent with LGRB correlations and if they have their own correlations among $E_{\rm p}$, $L_{\rm p}$ and $E_{\rm iso}$.
Recently, \citet{Zhang:2012} examined the \amati correlations for the 7 short and 105 long GRBs separately and confirmed quantitatively that they are significantly different from each other. On the other hand, concerning the \yonetoku correlation, they derived the correlation from the mixture of LGRBs and SGRBs and insisted, from a visual inspection, that SGRBs are consistent with their LGRB correlation. In fact, to argue the consistency between LGRBs and SGRBs, they should derive the correlations separately and compare them, as we will do below. Comparison of our results and \citet{Zhang:2012} will be given in section 4.

Table 1 shows our list of SGRB candidates which are selected as GRBs with
$T_{90}^{\rm rest} = T_{90}/(1+z) < 2~{\rm s}$ following \citet{Gruber:2011},
rather than $T_{90} < 2~{\rm s}$. The list contains the redshift $z$,
the rest frame duration $T_{90}^{\rm rest}$, the spectral peak energy $E_{\rm p}$,
the peak luminosity $L_{\rm p}$ in 64 ms of the observer-frame time bin,
the isotropic energy $E_{\rm iso}$, class of SGRB candidates which will be
explained later, and the reference. To make Table 1, we collected all GRBs
by the end of 2011 with the value of $T_{90}^{\rm rest} < 2s$,
the measured redshift $z$, the spectral peak energy $E_ {\rm p}$,
the peak flux $F_{\rm p, obs}$ and the fluence $S_{\rm obs}$ within
the energy range between $E_{\rm min}$ and $E_{\rm max}$ of each instrument. 
In order to obtain tighter correlations, the time bin of $F_{\rm p}$,
and then $L_{\rm p}$, should be defined in the time in GRB rest frame
as discussed in \cite{Tsutsui:2011,Tsutsui:2012a} for LGRBs.
However, the number of SGRBs is so small to determine the best time bin of
$L_{\rm p}$ that we simply adopt here 64 msec in the observer frame for
all SGRBs candidates.

For GRBs detected by Fermi/GBM (090423, 090510, 100117A, 100206, 100816A), 
we analyze the spectrum with the software package 
RMFIT\footnote{http://fermi.gsfc.nasa.gov/ssc/data/analysis/} 
(version 3.3rc8) and the GBM Response Matrices v1.8, 
 following the guidance of the RMFIT
tutorial\footnote{http://fermi.gsfc.nasa.gov/ssc/data/analysis/user/vc\_rmfittutorial.pdf}.  
For the other GRBs, we obtained the data from the reference in  1.
From these spectral parameters, peak fluxes and fluences, we can 
calculate the bolometric isotropic energy $E_{\rm iso}$ and the peak luminosity 
$L_ {\rm p}$ between the energy range of 1--100,000 keV in GRB rest frame 
using the Band function \citep{Band:1993}. 
Although in most of previous works,  $L_ {\rm p}$ and $E_{\rm iso}$ between 1--10,000 keV were adopted,  
in this paper we adopt 1--100,000 keV range, because 090510 has $E_{\rm p}\sim 8,000$ keV. 
$L_{\rm p}$ between 1--100,000 keV of GRB 090510 is 5 times larger than that of between 1-10,000 keV.
For 090424, 050709, 051221, 061006, 070714B, 071020, 080913, 100117A and 101219A, we used fixed high
energy photon index as $\beta = -2.25$, because we can not obtain high
energy photon index due to the lack of  number of photons.
For short GRBs with extended emission, $E_{\rm p}$ and $E_{\rm iso}$ were estimated for initial short/hard spikes.

Here we defined SGRB candidates as GRBs with $T_{90}^{\rm rest} < 2~{\rm s}$.
These are "candidates" because there might be some contamination from LGRBs
with relatively short duration \citep{Zhang:2009,Levesque:2010,Lu:2010}. 
\citet{Zhang:2009} proposed multiple observational criteria from their physical motivations, 
such as supernova (SN) association, specific star formation rate (SFR) of the host galaxy,
the location offset from the host galaxy, the duration, the hardness and the spectral lag, etc.  
However, because most of these observational properties are not available in many cases,
these criteria are not so useful in practice. In this study, we adopt much simpler
criteria by \cite{Lu:2010} which utilize the $E_{\rm p}$--$E_{\rm iso}$ correlation for LGRBs
as a discriminator against SGRBs. Thus, we define GRBs which have
$T_{\rm 90}^{\rm rest} < 2~{\rm s}$ and are consistent with $E_{\rm p}$--$E_{\rm iso}$ correlation
for LGRBs within 3-$ \sigma_{\rm int}$ dispersion level as "misguided SGRBs" and the others as "secure SGRBs".
That is, if a SGRB candidate is not consistent with the $E_{\rm p}$--$E_{\rm iso}$ correlation for LGRBs,
we regard it as a secure SGRB. 
In Table 1, we can find that misguided SGRBs tend to have longer $T_{90}^{\rm rest}$ and redshift than 
secure SGRBs. It might be not surprising because, the higher the redshift is, the more difficult it becomes to observe 
the long tail of the prompt emission. Then it is inevitable to underestimate $T_{90}^{\rm rest}$.
We should note that all SGRBs with extended emission in Table 1 (061006, 070714B, 101219A) belong 
to secure SGRBs, so the extended emission might be  a good indicator of secure SGRBs.

\begin{figure*}
\rotatebox{0}{\includegraphics[width=75mm]{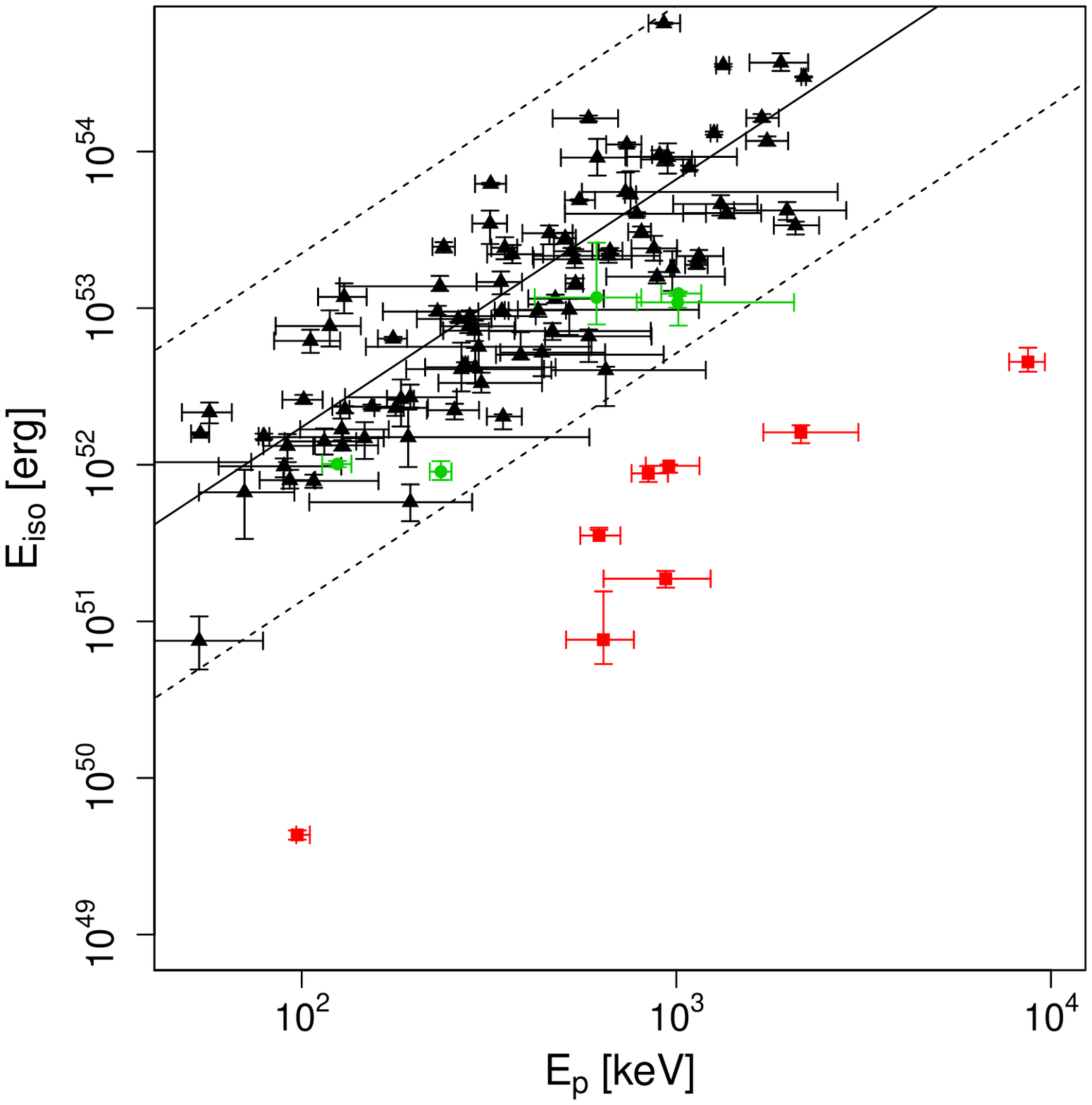}}
\rotatebox{0}{\includegraphics[width=75mm]{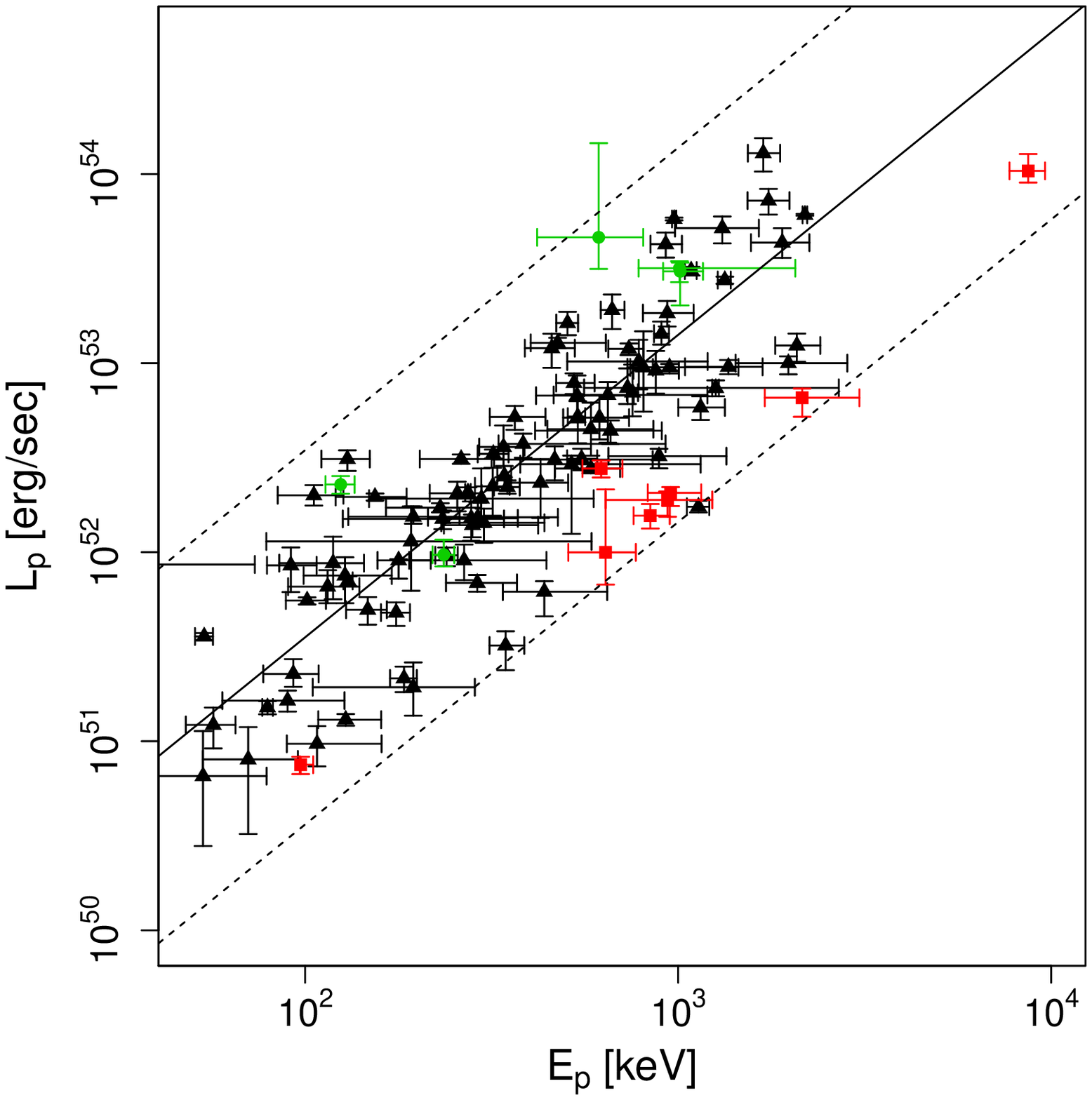}}
\caption{ The $E_ {\rm p}$ -- $E_{\rm iso}$  (left) and $E_{\rm p}$ -- $L_{\rm p}$ (right) diagrams.
The  LGRBs from \citet{Yonetoku:2010} are marked with black filled triangles, misguided SGRBs with green filled circles, 
and secure SGRBs with red filled squares. The best fit function and 3-$\sigma_{\rm int}$ dispersion of the correlations of LGRBs from \citet{Yonetoku:2010}  
are indicated with black solid and dotted lines, respectively. In this figure, the peak luminosities of LGRBs are defined by 1024 msec bin 
in observer frame, while those of SGRBs are by 64 msec bin in observer frame. 
 }
\label{Ep-Eiso-Lp-long}
\end{figure*}

Figure. \ref{Ep-Eiso-Lp-long} shows the $E_{\rm p}$--$E_{\rm iso}$ (left) and $E_{\rm p}$--$L_{\rm p}$ (right) diagrams
for both SGRB candidates in this Letter and LGRBs from \citet{Yonetoku:2010}.
In the left of Fig. 1, the best fit function and 3-$\sigma_{\rm int}$ dispersion region of $E_{\rm p}$--$E_{\rm iso}$ correlation
for LGRBs  are indicated by the  black solid and dotted lines, respectively.
A misguided SGRB which is located within 3-$\sigma_{\rm int}$ dispersion region of the $E_{\rm p}$--$E_{\rm iso}$ correlation
for LGRBs is marked by a green filled circle, while  a secure SGRB by a red filled square. 
We can see that  the secure SGRBs are always under the best fit function of $E_{\rm p}$--$E_{\rm iso}$
correlation for LGRBs although it can be above it from our definition of the secure SGRB.
This suggests that $E_{\rm p}$--$E_{\rm iso}$  correlation might exist even for secure SGRBs.
Similar argument was already discussed in previous studies \citep{Amati:2006,Ghirlanda:2009,Zhang:2012}.
We estimate the best fit function of $E_{\rm p}$--$E_{\rm iso}$ relation for secure SGRBs 
and quantitatively check these previous arguments.

\begin{figure*}
\rotatebox{0}{\includegraphics[width=75mm]{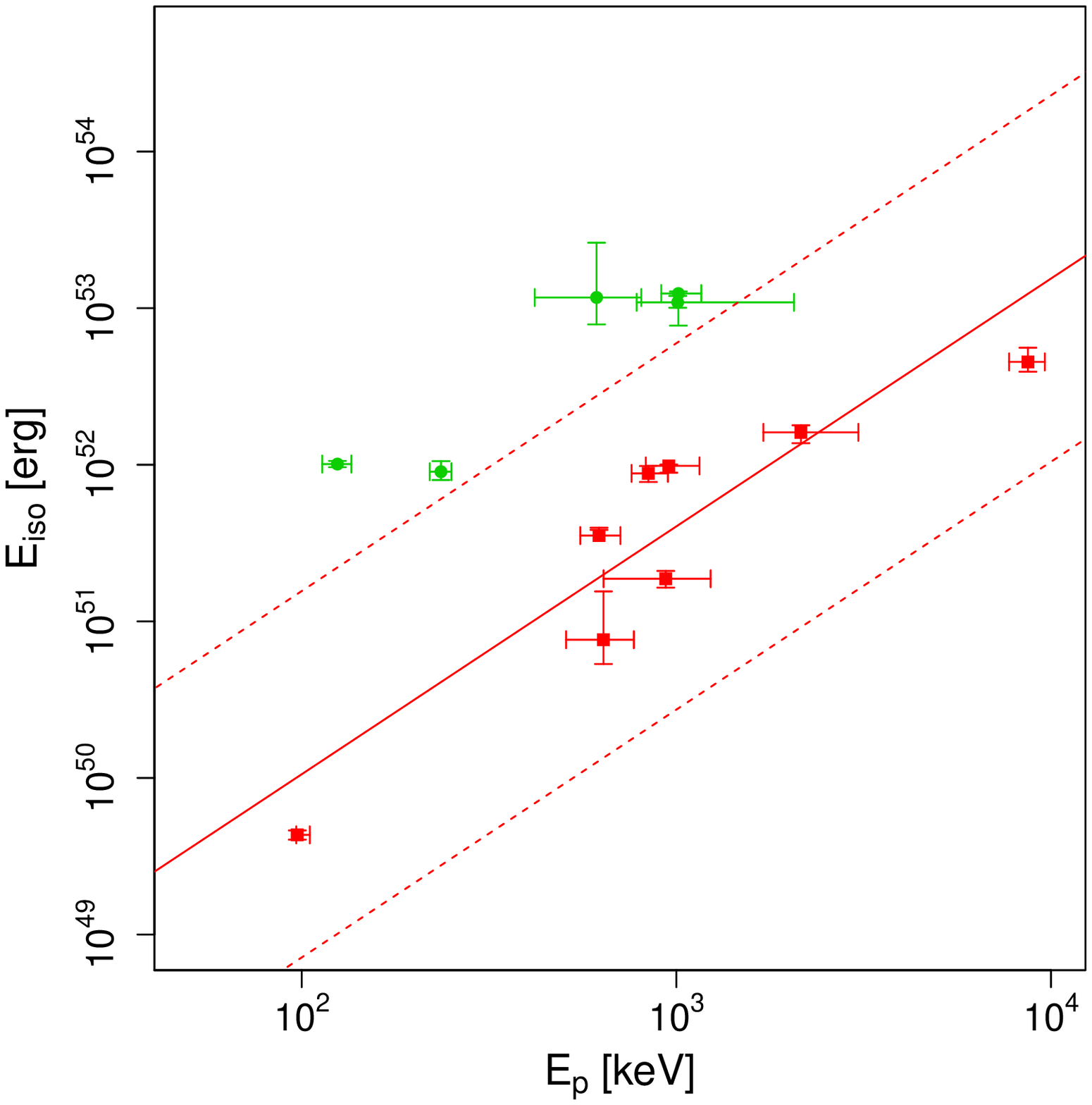}}
\rotatebox{0}{\includegraphics[width=75mm]{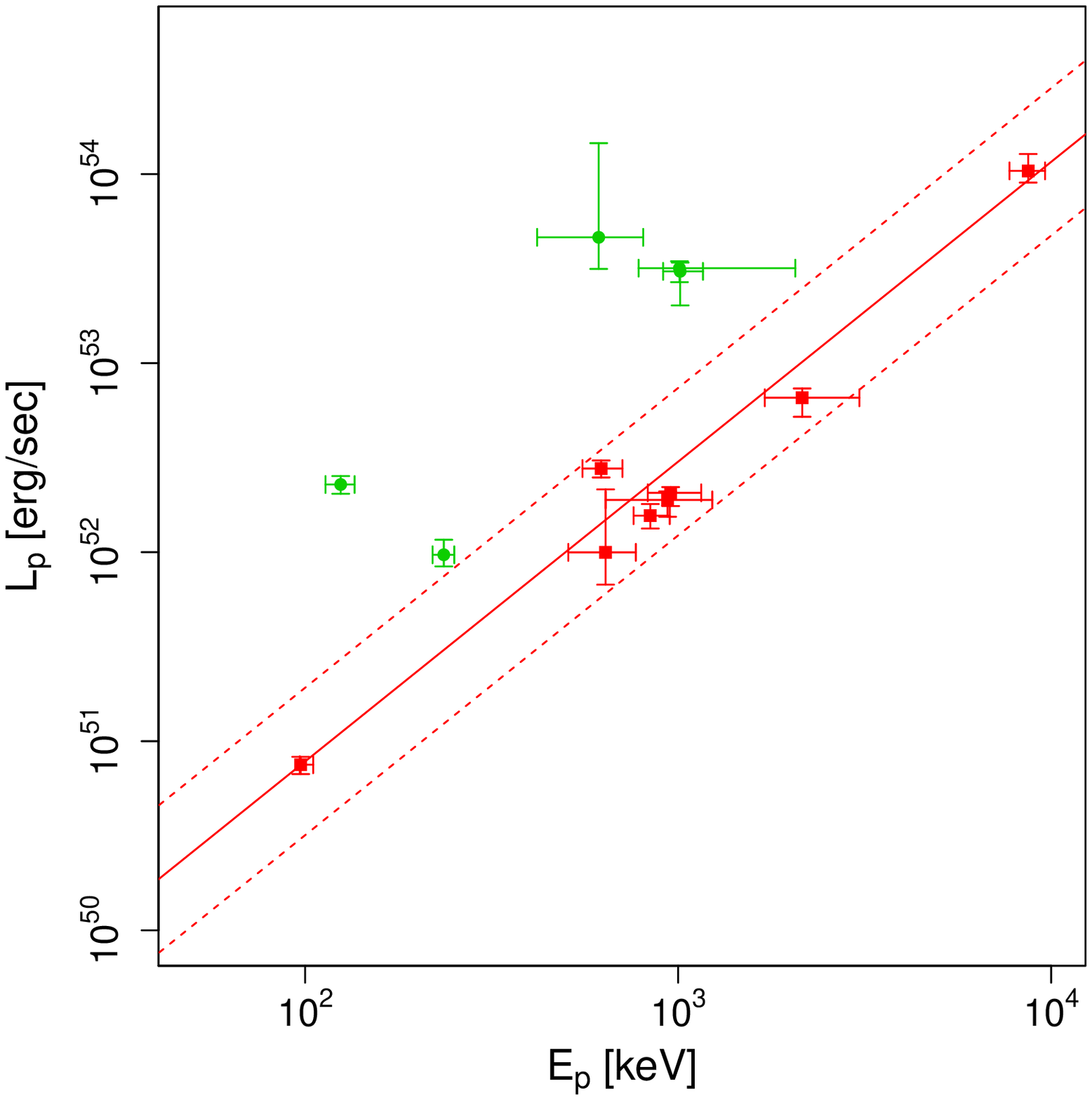}}
\caption{(Left) The $E_ {\rm p}$ -- $E_{\rm iso}$ diagram for SGRBs.
 (Right) The $E_ {\rm p}$ -- $L_ {\rm p}$ diagram for SGRBs.
In each figure, misguided SGRBs are marked with green filled circles, 
and secure SGRBs with red filled squares. The best fit function and 3-$\sigma_{\rm int}$ dispersion are indicated with 
red solid and dotted lines, respectively.
 }
\label{Ep-Lp-Eiso}
\end{figure*}

Let us  assume a linear correlation in logarithm as$\log E_{\rm iso}=A+B ( \log E_{\rm p}-<\log E_{\rm p}> )$, 
where angle bracket denotes an average, 
 and a chi square function as  
 \begin{equation}
 \chi^2(A,B)=\sum\frac{(\log E_{\rm iso}^i -A-B\log (E^i_{\rm p}/774.5 {\rm keV}))^2}{\sigma_{E_{\rm iso}, i}^2+B^2\sigma_{E_{\rm p}, i}^2+\sigma_{\rm int}^2},
 \end{equation}
where $\sigma_{E_{\rm iso}}$ ($\sigma_{E_{\rm p}}$) is statistical error of $E_{\rm iso}$ ($E_{\rm p}$) for each burst 
, and $\sigma_{\rm int}$ (the same for all bursts) is the intrinsic dispersion of the relation added as extra component of $E_{\rm iso}$ scatter, 
since statistical errors of  $\sigma_{E_{\rm iso}}$ and $\sigma_{E_{\rm p}}$ do not account for a large scatter of the relation. 
In this paper, we assume the intrinsic distribution around a relation  is gaussian and we estimate a value of $\sigma_{\rm int}$ as the 
value which makes a reduced chi square value unity by trial and error.
Then, the best fitted values and 1-$\sigma$ errors of $A$ and $B$ are estimated with the fixed value of $\sigma_{\rm int}$. 
In the left of Fig. 2 we plot only secure (red filled square) and misguided (green filled circle) 
SGRBs in $E_{\rm p}$--$E_{\rm iso}$ diagram. The red solid line is
the best fit fuction of $E_{\rm p}$--$E_{\rm iso}$ correlation for secure SGRBs given by 
\begin{equation}
E_{\rm iso} =10^{51.42 \pm 0.15}{\rm erg}
(\frac{E_{\rm p}}{\rm 774.5~keV})^{1.58 \pm 0.28}. 
\end{equation}
The logarithmic correlation coefficient($r$) is 0.91 with the chance probability($p$) of $1.5\times 10^{-3}$ and
$\sigma_{\rm int}=0.39$. The dotted red line shows the 3-$\sigma_{\rm int}$ dispersion. 
We can say that $E_{\rm p}$--$E_{\rm iso}$ correlation exists for secure SGRBs also.
Therefore although it is correct that SGRBs do not obey $E_{\rm p}$--$E_{\rm iso}$ correlation for LGRBs, which has been claimed, 
they do obey the different $E_{\rm p}$--$E_{\rm iso}$ correlation
with almost the same power law index but   a factor $\sim 100$ smaller amplitude in $E_{\rm iso}$.

Now let us discuss $E_{\rm p}$--$L_{\rm p}$ correlation. In the right of
Figure \ref{Ep-Eiso-Lp-long}, we plot secure (filled square) 
and misguided (green filled circle) SGRBs in $E_{\rm p}$--$L_{\rm p}$ diagram. The best-fit function
and 3-$\sigma_{\rm int}$ dispersion region of $E_{\rm p}$--$L_{\rm p}$ correlation for LGRBs from \citet{Yonetoku:2010}
are indicated by the  black solid and dotted lines, respectively. 
All of secure and misguided SGRB lie inside the 3-$\sigma_{\rm int}$ dispersion region of $E_{\rm p}$--$L_{\rm p}$ 
correlation for LGRBs. This might lead us an argument that SGRBs follows  the same $E_{\rm p}$--$L_{\rm p}$ correlation for LGRBs 
as discussed in previous studies \citep{Ghirlanda:2009,Zhang:2012}.
However if we focus only on secure SGRBs,  all of them are located under the black solid line in the right of Fig. 1. 
This result seems to be unnatural because if they really come from the same $E_{\rm p}$--$L_{\rm p}$ correlation for LGRBs, 
about half of them must be above the black solid line.
This fact implies the existence of $E_{\rm p}$--$L_{\rm p}$ correlation for secure SGRBs
although the best fit function of them is within the 3-$\sigma_{\rm int}$ dispersion of
$E_{\rm p}$--$L_{\rm p}$ correlation for LGRBs. 
In  this paper, we obtain a $E_{\rm p}$--$L_{\rm p}$ correlation only from secure SGRBs for the  first time 
and compare it with the relation for LGRBs.

In the right of Figure 2 we plot only
secure (filled square) and misguided (green filled circle) SGRBs
in $E_{\rm p}$--$L_{\rm p}$ diagram. It is clear that secure SGRBs have their own correlation
and the best-fit function is given by,
\begin{equation}
L_{\rm p} = 10^{52.29 \pm 0.066}{\rm erg~ s^{-1}}
(\frac{E_{\rm p}}{\rm 774.5~keV})^{1.59 \pm 0.11} 
\label{eq:yonetoku}
\end{equation}
with $r=0.98$, $p=1.5 \times 10^{-5}$ and $\sigma_{\rm int}=0.13$.
The dotted red line shows the 3-$\sigma_{\rm int}$ dispersion. 
From the value of $r$ amd $p$, we can say
that the $E_{\rm p}$--$L_{\rm p}$ correlation for secure SGRBs is tighter than the $E_{\rm p}$--$E_{\rm iso}$
correlation for SGRBs. For this reason we use Eq. (\ref{eq:yonetoku}) as a distance indicator
in chapter \ref{redshift} to determine the pseudo redshift of SGRBs without measured redshift. 

The best fit function for long GRBs in \citet{Yonetoku:2010} is given by,
\begin{equation}
L_{\rm p} = 10^{52.97}{\rm erg~ s^{-1}}
(\frac{E_{\rm p}}{\rm 774.5~keV})^{1.60 }.
\label{eq:yonetoku2010}
\end{equation}
Comparing equation (3) and (4), we can say that SGRBs obey $E_{\rm p}$--$L_{\rm p}$ correlation with almost the same power law index but a factor $\sim$ 5
smaller amplitude in $L_{\rm p}$ with 10-$\sigma$ statistical significance. 
Here we compared the \yonetoku correlation for SGRBs of our sample with the one for LGRBs of \citet{Yonetoku:2010}, 
while \citet{Zhang:2012} did the one for SGRBs of their sample with the one for LGRBs of \citet{Ghirlanda:2010}.
One may suspect that the difference might come from the difference of the sample both of long and short GRBs, 
but we will show that we can get the same result with this paper even if we use \citet{Ghirlanda:2010} and \citet{Zhang:2012} sample in 
section 4.

\begin{table*}
\begin{minipage}{\textwidth}
\caption{List of  all SGRB candidates until the end of 2011 used for the analysis. Each column corresponds
to  the  redshift $z$, the rest frame duration $T_{90}^{\rm rest}=T_{90}/(1+z)$,  the  spectral peak energy
$E_ {\rm p}$, the peak luminosity $L_ {\rm p}$ in  64 ms   of the  observer frame time bin,  the isotropic energy
$E_{\rm iso}$, class of SGRB candidates and the reference, respectively. For details see the text.}
\begin{tabular}{lccccccc}																										
\hline\hline																											
GRB	&	redshift	&	$T_{\rm 90}^{\rm rest}$	[sec] &	$E_{\rm p}$	[keV] &					$L_{\rm p}$					[erg/s] &			$E_{\rm iso}$	[erg] &			class		& ref\footnote{References for spectral parameters, peak fluxes and fluences: (1)\citet{GCN2754} ; (2) \citet{Villasenor:2005} ; (3) \citet{GCN4394,GCN4388} ; 
(4) \citet{GCN5710} ; (5) \citet{GCN6638,GCN6637} ; (6) \citet{ GCN6960} ; (7) \citet{GCN8256,GCN8222} ; (8) This work ; (9) \citet{GCN11470}.} 
\\ \hline
040924	&	0.86	&	0.81	& $	124.55	^{+	11.15	}_{	-11.15	}$ & $ (	2.28	^{+	0.25	}_{	-0.24	} ) \times 10^{	52	} $ & $ (	1.01	^{+	0.05	}_{	-0.05	} ) \times 10^{	52	} $ &	misguided	& (1) \\
050709\footnote{70 msec peak luminosity}	&	0.16	&	0.60	& $	97.32	^{+	7.76	}_{	-0.58	}$ & $ (	7.51	^{+	0.76	}_{	-0.81	} ) \times 10^{	50	} $ & $ (	4.33	^{+	0.29	}_{	-0.30	} ) \times 10^{	49	} $ &	secure& (2)	\\
051221	&	0.55	&	0.91	& $	621.69	^{+	87.42	}_{	-67.69	}$ & $ (	2.77	^{+	0.29	}_{	-0.29	} ) \times 10^{	52	} $ & $ (	3.53	^{+	0.43	}_{	0.31	} ) \times 10^{	51	} $ &	secure& (3)	\\
061006	&	0.44	&	0.35	& $	954.63	^{+	198.39	}_{	-125.86	}$ & $ (	2.06	^{+	0.15	}_{	-0.31	} ) \times 10^{	52	} $ & $ (	9.83	^{+	0.20	}_{	-0.94	} ) \times 10^{	51	} $ &	secure& (4)	\\
070714B	&	0.92	&	1.04	& $	2150.40	^{+	910.39	}_{	-443.52	}$ & $ (	6.56	^{+	0.79	}_{	-1.36	} ) \times 10^{	52	} $ & $ (	1.61	^{+	0.18	}_{	-0.24	} ) \times 10^{	52	} $ &	secure & (5)	\\
071020	&	2.15	&	1.11	& $	1012.69	^{+	152.94	}_{	-101.33	}$ & $ (	3.06	^{+	0.35	}_{	-1.04	} ) \times 10^{	53	} $ & $ (	1.24	^{+	0.04	}_{	-0.47	} ) \times 10^{	53	} $ &	misguided& (6)	\\
080913	&	6.70	&	1.04	& $	1008.05	^{+	1052.52	}_{	-224.54	}$ & $ (	3.18	^{+	0.28	}_{	-0.50	} ) \times 10^{	53	} $ & $ (	1.09	^{+	0.11	}_{	-0.08	} ) \times 10^{	53	} $ &	misguided& (7)	\\
090423	&	8.26	&	1.30	& $	612.36	^{+	193.53	}_{	-193.53	}$ & $ (	4.63	^{+	9.95	}_{	-1.48	} ) \times 10^{	53	} $ & $ (	1.17	^{+	1.45	}_{	-0.38	} ) \times 10^{	53	} $ &	misguided & (8)\\
090510	&	0.90	&	0.16	& $	8679.58	^{+	947.69	}_{	-947.69	}$ & $ (	1.04	^{+	0.24	}_{	-0.14	} ) \times 10^{	54	} $ & $ (	4.54	^{+	1.05	}_{	-0.61	} ) \times 10^{	52	} $ &	secure & (8)	\\
100117A	&	0.92	&	0.16	& $	936.96	^{+	297.60	}_{	-297.60	}$ & $ (	1.89	^{+	0.21	}_{	-0.35	} ) \times 10^{	52	} $ & $ (	1.87	^{+	0.23	}_{	-0.23	} ) \times 10^{	51	} $ &	secure & (8)	\\
100206	&	0.41	&	0.09	& $	638.98	^{+	131.21	}_{	-131.21	}$ & $ (	9.98	^{+	11.50	}_{	-3.25	} ) \times 10^{	51	} $ & $ (	7.63	^{+	7.89	}_{	-2.29	} ) \times 10^{	50	} $ &	secure & (8)	\\
100816A	&	0.81	&	1.11	& $	235.36	^{+	15.74	}_{	-15.74	}$ & $ (	9.69	^{+	1.95	}_{	-1.28	} ) \times 10^{	51	} $ & $ (	9.03	^{+	1.52	}_{	-1.04	} ) \times 10^{	51	} $ &	misguided & (8)	\\
101219A	&	0.72	&	0.35	& $	841.82	^{+	107.56	}_{	-82.50	}$ & $ (	1.56	^{+	0.24	}_{	-0.23	} ) \times 10^{	52	} $ & $ (	8.81	^{+	1.00	}_{	-1.05	} ) \times 10^{	51	} $ &	secure &(9) \\ \hline
\end{tabular}
\end{minipage}
\end{table*}

\section{Redshift Estimation}
\label{redshift}
From the analysis in the previous section, the $E_{\rm p}$--$L_{\rm p}$ correlation
for SGRBs derived would be a better distance
indicator of SGRBs than the $E_{\rm p}$ -- $E_{\rm iso}$ correlation.
The best-fit function of Eq.~(\ref{eq:yonetoku}) can be rewritten
using the observed quantities as
\begin{equation}
\label{eq:distance}
\frac{d_{L}^{2}}{(1+z)^{1.59}} =
\frac{10^{52.29}{\rm erg~ s^{-1}}}{4 \pi F_{p}} 
( \frac{E_{p,obs}}{774.5~{\rm keV}} )^{1.59},
\end{equation}
where $d_{L}$, $E_{p,obs}$ and $F_{p}$ are the luminosity distance,
the peak energy at the observer's rest frame and 
the peak flux, respectively. The right hand side of this equation
consists of only the observable quantities. Therefore assuming
the $\Lambda$-CDM cosmology with $(\Omega_m, \Omega_{\Lambda}) = (0.3,0.7)$,
we can uniquely determine the redshift through the luminosity distance
which is a function of redshift. We call this as the pseudo redshift.
The important point here is that the left hand side of Eq. (3) is a 
monotonically increasing function of $z$ from zero for $z=0$ to $\infty$
for $z=\infty$ so that a unique solution exists for any observed value
of the right hand side. 
To estimate uncertainties of pseudo redshifts, 1-$\sigma$ intrinsic dispersion of 
the relation on the normalization of the equation (\ref{eq:yonetoku}) are taken into account.

\begin{figure}
\rotatebox{0}{\includegraphics[width=80mm]{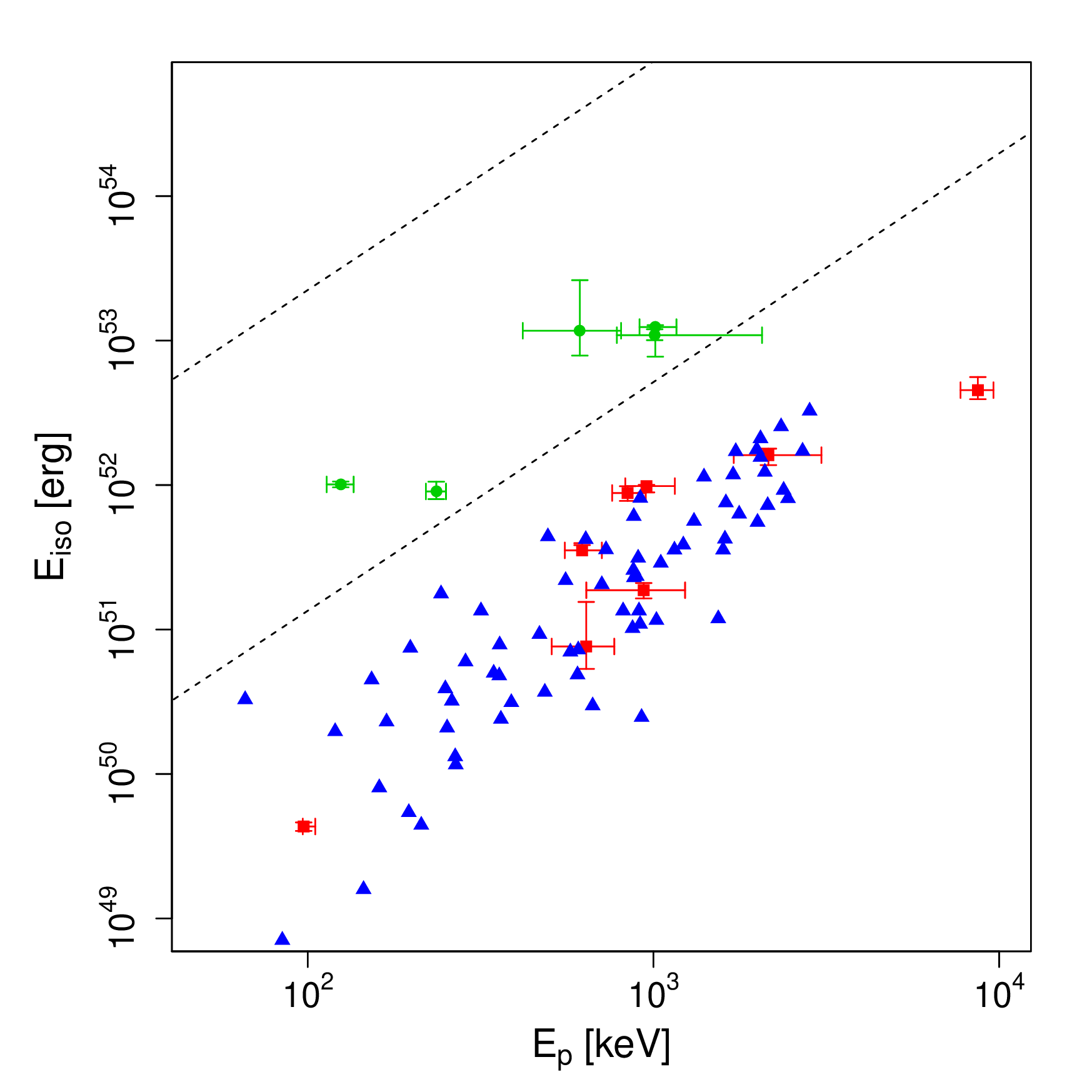}}
\caption{The pseudo $E_{\rm p}$--$E_{\rm iso}$ diagram from 71 BATSE SGRBs (blue filled triangles).
The secure and misguided SGRBs are marked with red filled squares and green filled triangles, respectively.
The 3-$\sigma_{\rm int}$ dispersion region of LGRBs is indicated by the dotted lines.}
\label{pseudo-Ep-Eiso}
\end{figure}

\begin{figure}
\rotatebox{0}{\includegraphics[width=75mm]{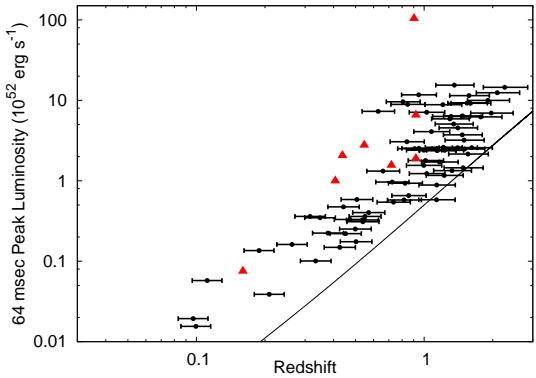}}
\caption{The redshift distribution of SGRBs estimated by 
the best fit $E_{\rm p}$--$L_{\rm p}$  correlation for SGRBs(Eq.(2)). 
We used 71 bright BATSE SGRBs reported by \citet{Ghirlanda:2009}, 
and succeeded in estimating the redshift for all events.
Black dots are the pseudo $z$ and $L_p$  while red filled squares are those of 
 secure SGRBs. Redshift $z$ ranges from $0.097$ to  $2.581$ with the mean $<z>$ of 1.05. 
Note that for $\it Swift$ LGRBs $<z>\sim 2.16$ \citep{Jakobsson:2012}.
The solid line is a reference 
of flux limit of $F_{p} = 10^{-6}~{\rm erg~cm^{-2}s^{-1}}$.}
\label{z-Lp}
\end{figure}

We used the data of 79 bright SGRBs observed by CGRO-BATSE 
reported by \citet{Ghirlanda:2009}. The $E_{p,obs}$ values were 
not measured for 8~samples, so that finally we use 71~samples listed
in their list. They selected the events with the burst duration of 
$T_{90, obs} < 2~{\rm sec}$ and the peak photon flux of
$P > 3~{\rm photons~cm^{-2}s^{-1}}$ in 64~msec time resolution.
They basically used the cutoff power-law (CPL) model to 
measure the spectral parameters. 
Using Eq.~(\ref{eq:distance}), we can estimate the pseudo
redshifts of all 71 SGRBs. In Fig.~\ref{z-Lp}, we show the
distribution on the $(z, L_p)$ plane. The solid line is a reference 
of flux limit of $F_{p} = 10^{-6}~{\rm erg~cm^{-2}s^{-1}}$. 
We found that the pseudo $z$ ranges from $0.097$ to  $2.581$. The mean pseudo redshift $<z>$  is 1.05. 
However, We expect more dim  SGRBs under  the solid line.
We note here that for $\it Swift$ LGRBs  $<z>=2.16$ \citep{Jakobsson:2012}. 
In Fig. \ref{pseudo-Ep-Eiso},
there are few SGRBs for low $z$ with large $L_ {\rm p}$.  This might be  a selection effect since the comoving volume is 
in proportion to
$z^3$ for $z < 1$ so that the SGRB with large $L_ {\rm p}$ would be rare. For $z > 1$, we do not see such an effect. 
Although in principle we can
determine the luminosity function as in \citet{Yonetoku:2004}, in practice, the number of SGRBs is too small to do so.
In Figure \ref{pseudo-Ep-Eiso}, we plot pseudo $E_{\rm p}$--$E_{\rm iso}$ diagram for
71 BATSE bright SGRBs (blue filled triangle) with secure and misguided SGRBs in the left of Figure \ref{Ep-Lp-Eiso}.
The distribution of BATSE bright SGRBs is very similar to that of the secure SGRBs.
Furthermore, all of them are out of the 3-$\sigma_{\rm int}$ region of LGRBs indicated by the dotted lines
so that there is no misguided SGRBs in them.

Figure. \ref{z-Lp} shows that the pseudo redshift distribution of the bright BATSE SGRBs has a rather sharp cut
off around $z=2.5$. This favors the compact star merger scenario of SGRBs since the time is needed for the binary
to merge so that there might be the maximum redshift of SGRBs. 
Our result seems to be different from the result of \citet{Ghirlanda:2004} in which the distribution of pseudo redshifts of 
SGRBs are similar to that of LGRBs.
This is because they assumed the $E_{\rm p}$--$L_{\rm p}$ correlation for LGRBs in \citet{Yonetoku:2004}.
The correlation for LGRBs can be rewritten as $L_{\rm p}=10^{53.15}(E_{\rm p}/774.5)^2$ which is 
$10^{0.9}$ times brighter than Eq. (2) and then assuming such a bright correlation overestimates pseudo redshifts. 
We used the correlation constructed with only secure SGRBs, and then our result would be much more reliable.

\section{Discussions}

As mentioned before, the comparison of the distribution of LGRBs and SGRBs in $E_{\rm p}$--$E_{\rm iso}$ and $E_{\rm p}$--$L_{\rm p}$ planes was performed in \citet{Zhang:2012}. As to the former, they recognized the difference in the distribution and found that the $E_{\rm p}$--$E_{\rm iso}$ correlation from SGRBs is almost parallel but dimmer by a factor of 10 compared with the one from LGRBs. This is reasonably consistent with our result. However, as to the latter, they insisted that SGRBs follow the same correlation as the one derived from LGRBs, which is in contradiction with our analysis. Below, we will discuss the origin of this discrepancy.

\begin{figure*}
\rotatebox{0}{\includegraphics[width=75mm]{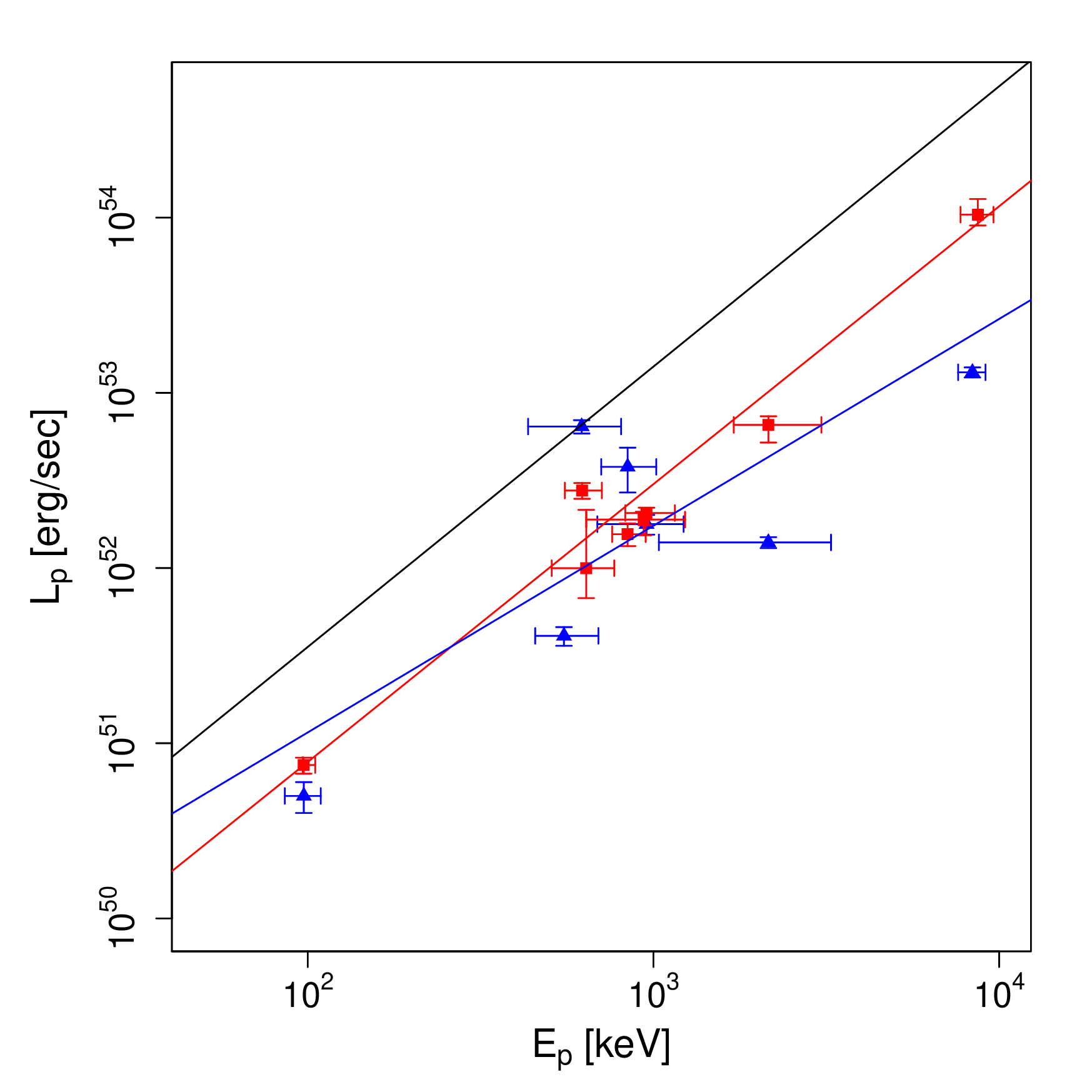}}
\rotatebox{0}{\includegraphics[width=75mm]{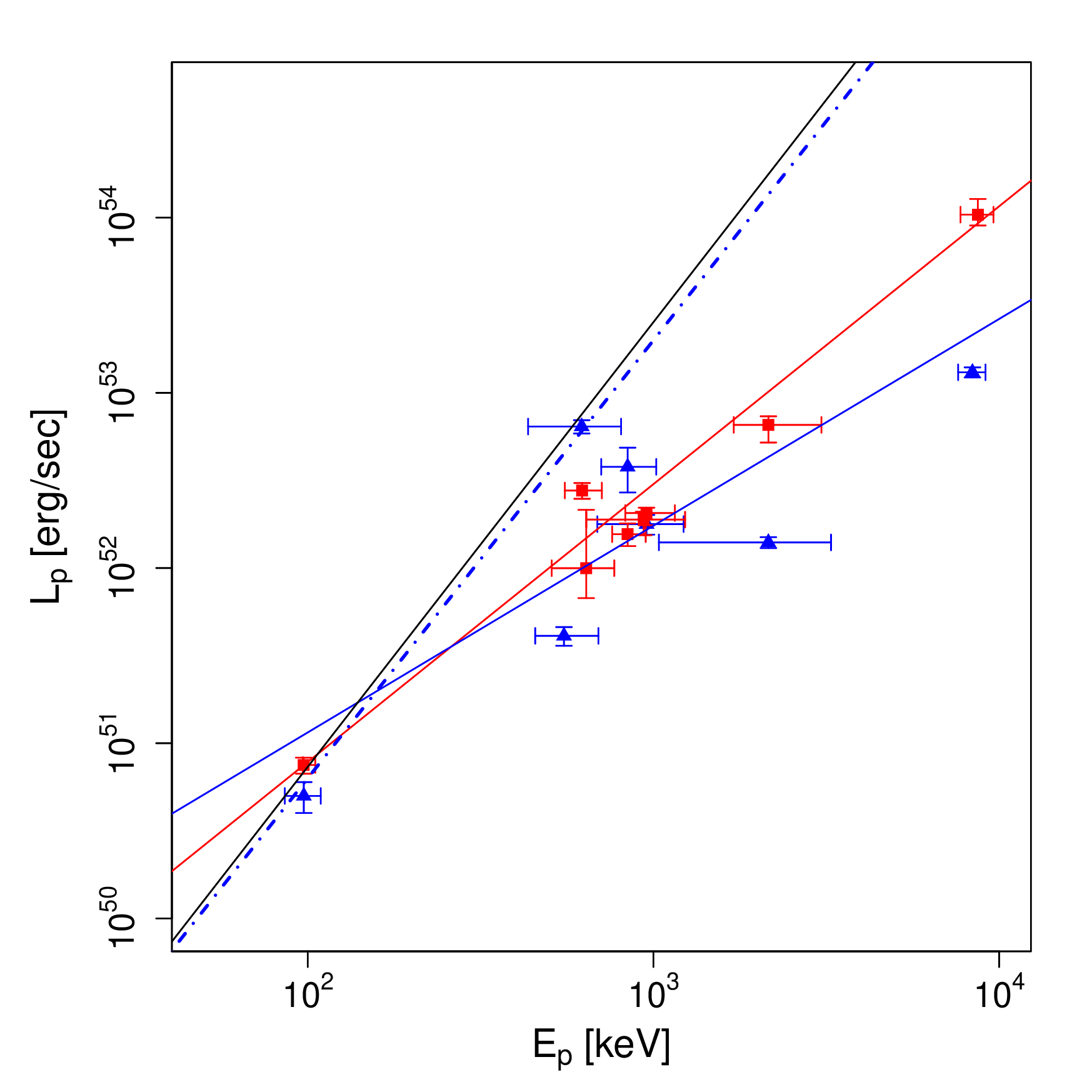}}
\caption{(Left) The $E_{\rm p}$--$L_{\rm p}$ diagram for SGRBs. Our eight secure SGRBs are marked with red filled squares and seven SGRBs from Zhang et al. (2012) with blue filled triangles. The best fit for each sample are indicated with red solid line and blue solid line, respectively and the black line is the best fit for LGRBs from \citet{Yonetoku:2010}. (Right) The same diagram as the left panel, but the black line is the the best fit for LGRBs from \citet{Ghirlanda:2010} and blue dash-doted line is the one for combined LGRBs and SGRBs from \citet{Zhang:2012}.
}
\label{vszhang12}
\end{figure*}

The left panel of Figure \ref{vszhang12} shows \yonetoku diagram for our secure SGRB sample (red filled circles) and SGRB sample from \citet{Zhang:2012} (blue filled squares) with the best-fit line for each sample. Here it should be noted that the best-fit line solely from SGRBs was not derived in \citet{Zhang:2012} and was newly derived here. The best-fit line for LGRBs of \citet{Yonetoku:2010} is also plotted with a black solid line. We can see that all but one events are located below the LGRB line. This fact indicates that SGRBs are systematically dimmer than LGRBs with the same $E_{\rm p}$ even if we consider SGRB sample by \citet{Zhang:2012}.

Here we should note that 7 of 8 SGRBs in our sample are actually the same events with \citet{Zhang:2012}, though they have different values of $L_{\rm p}$ which leads to the different best-fit lines. Let us comment on the difference in each event. First of all, our peak luminosities are uniformly defined by the 64-msec time resolution in obesrver frame for all SGRBs, while \citet{Zhang:2012} used different time resolutions (from 4 ms to 1024 ms). This is because they adopted the values of $L_{\rm p}$ reported by multiple observation teams. The value of 4-msec peak luminosity is typically a few times larger than that of 1024-msec peak luminosity \citep{Tsutsui:2011,Tsutsui:2012a} so that using different time resolution to define $L_{\rm p}$ would make artificial dispersion in the $E_{\rm p}$--$L_{\rm p}$ correlation \citep{Yonetoku:2010}. Secondly, we integrate energy spectra between 1-100,000 keV in GRB rest frame to calculate $L_{\rm p}$, while an energy range of 1-10,000 keV was considered in \citet{Zhang:2012}. Therefore they tend to underestimate $L_{\rm p}$ compared with us. From these reasons, the values of $L_{\rm p}$ are different in the two samples and, we believe, our sample is more reliable compared with \citet{Zhang:2012}.

On the other hand, the correlation for the LGRBs are also different between our analysis and \citet{Zhang:2012}. The right panel of Figure \ref{vszhang12} is the same diagram as the left panel, but the best-fit line is for LGRBs in \citet{Ghirlanda:2010} which \citet{Zhang:2012} uses. The best-fit line for combined short and long GRB sample obtained by \citet{Zhang:2012} is also plotted with a blue dash-doted line. Although the best-fit lines are significantly different from ours, the same tendency can still be seen.

It is beyond the scope of this paper to explain the difference between the \yonetoku correlations for LGRBs from \citet{Ghirlanda:2010} and the one from \citet{Yonetoku:2010}, and we just make a short remark on this. The major difference comes from the treatment of GRB 060218. The former regarded it as a ordinary LGRB, while in the latter it was regarded as an outlier by a statistical argument. Because GRB 060218 is located far away from the \yonetoku correlation of \citet{Yonetoku:2010} (more than 8-$\sigma$), it makes the best-fit line much steeper like the one of \citet{Ghirlanda:2010}. Anyway, it seems to be robust that SGRBs have systimatically smaller $L_{\rm p}$ than LGRBs for a given $E_{\rm p}$, even if we consider the possible systematic errors in LGRBs, as well as SGRBs.

In this paper, we suggested possible correlations among $E_{\rm p}$, $L_{\rm p}$ and $E_{\rm iso}$  even for SGRBs. However, the correlations for SGRBs are much dimmer than those for LGRBs. The $E_{\rm p}$--$E_{\rm iso}$ ($E_{\rm p}$--$L_{\rm p}$) correlation for SGRBs is located approximately $10^2$ ($5$) times below the one for LGRBs. For the $E_{\rm p}$--$L_{\rm p}$ correlation for SGRBs, similar arguments have been made by some authors \citep{Ghirlanda:2009,Zhang:2012}, but we for the first time argue that there exist distinct $E_{\rm p}$--$E_{\rm iso}$ and $E_{\rm p}$--$L_{\rm p}$ correlations for SGRBs.

\begin{figure}
\rotatebox{0}{\includegraphics[width=85mm]{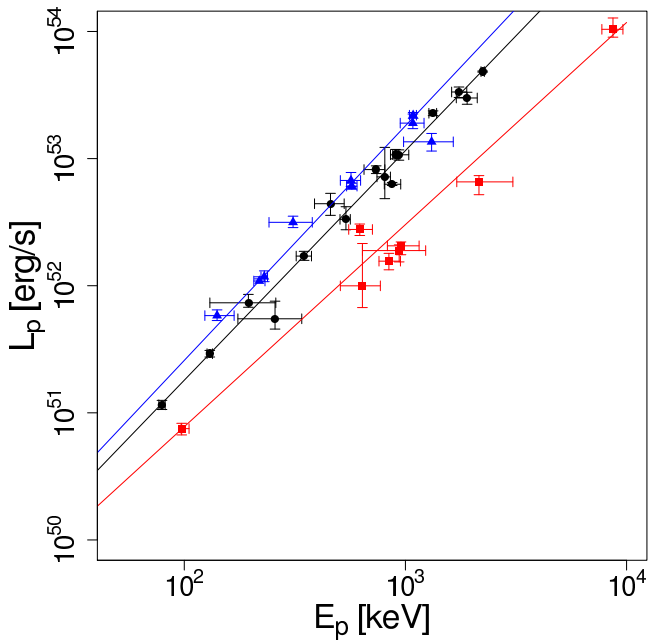}}
\caption{The $E_{\rm p}$ -- $L_{\rm p}$ diagram both for short and long GRBs. The LGRBs from gold sample in \citet{Tsutsui:2012a} are marked with black filled circles (small-$ADCL$ GRBs) and blue filled triangles (large-$ADCL$ GRBs). The outliers of gold sample in \citet{Tsutsui:2012a} and misguided GRBs are removed from this figure.
}
\label{Ep-Lp}
\end{figure}

The distinction between SGRBs and LGRBs becomes much clearer if we use the gold sample of LGRBs compiled by \citet{Tsutsui:2012a}. \citet{Tsutsui:2012a} argued that there are two $E_{\rm p}$--$L_{\rm p}$ correlations, one is for small-$ADCL$  GRBs and the other is for large-$ADCL$ GRBs, where $ADCL$ stands for Absolute Deviations from  Constant Luminosity. In figure \ref{Ep-Lp}, we shows the $E_{\rm p}$--$L_{\rm p}$ diagram for small-$ADCL$ LGRBs (black filled circles), large-$ADCL$ LGRBs (blue filled triangles), and secure SGRBs (red filled squares). The outliers of gold sample in \citet{Tsutsui:2012a} and misguided SGRBs are removed from this figure. We can see the existence of three distinct $E_{\rm p}$--$L_{\rm p}$ correlations with almost the same power law index and different amplitudes.

The accurate functional forms of $E_{\rm p}$--$E_{\rm iso}$ and $E_{\rm p}$--$L_{\rm p}$ correlation are very important to study the progenitor and the radiation mechanism of SGRBs. At present the intrinsic dispersion is rather large, that is , 0.13(0.39) in logarithm for $E_{\rm p}$--$L_{\rm p}$($E_{\rm p}$--$E_{\rm iso}$), respectively. This is mainly due to the small number of secure SGRBs, which prevents more detailed analysis. In conclusion we need more data of SGRBs with accurate $z$, $E_ {\rm p}$, $L_ {\rm p}$ and $E_{\rm iso}$ to confirm or refute the $E_{\rm p}$--$E_{\rm iso}$ and $E_{\rm p}$--$L_{\rm p}$ correlations for SGRBs suggested in this Letter.

\section*{Acknowledgments}

This work is supported in part by the Grant-in-Aid for Young Scientists (B) 
from the Japan Society for Promotion of Science (JSPS), No.24740116(RT),
by the Grant-in-Aid from the Ministry of Education, Culture, Sports, Science and Technology
(MEXT) of Japan, No.23540305, No. 24103006 (TN), No.20674002 (DY),
No.23740179, 24111710, and 24340048 (KT),
and by the Grant-in-Aid for the global COE program 
{\it The Next Generation of Physics, Spun from Universality 
and Emergence} at Kyoto University. 

\bibliographystyle{mn2e}
\bibliography{tsutsui}

\end{document}